\documentclass[twocolumn,aps,prl,showpacs,superscriptaddress]{revtex4}

\usepackage{graphicx}
\usepackage{amsmath,amssymb}
\usepackage{color}

\begin{document}


\title{Proposed experiment to exclude higher-than-quantum violations of the Bell inequality}


\author{Ad\'an Cabello}
\email{adan@us.es}
\affiliation{Departamento de F\'{\i}sica Aplicada II, Universidad de Sevilla, E-41012 Sevilla, Spain}


\date{\today}



\begin{abstract}
 We show that a recent observation by Yan leads to a method to {\em experimentally test} whether a higher-than-quantum violation of the Clauser-Horne-Shimony-Holt Bell inequality is possible (assuming that the sum of probabilities of pairwise exclusive propositions cannot exceed~1). The test requires reaching the maximum quantum violation of a noncontextuality inequality involving sequences of three compatible measurements on a five-dimensional quantum system.
\end{abstract}



\maketitle


{\em Introduction.---}For years, and since the work of Popescu and Rohrlich \cite{PR94}, one of the most important problems in quantum mechanics (QM) has been finding the answer to the question of what fundamental principle prevents higher violations of the simplest Bell inequality \cite{Bell64}. The simplest Bell inequality is the Clauser-Horne-Shimony-Holt (CHSH) inequality \cite{CHSH69} and its maximum quantum violation is the so-called Tsirelson's bound \cite{Cirelson80}.

Different solutions to this problem have been proposed recently. They range from an information-theoretic principle (namely, information causality) applied to a bipartite scenario \cite{PPKSWZ09}, through an observational principle (namely, macroscopic locality) suitably formalized \cite{NW09}, to the observation that the Tsirelson bound is determined by other properties of QM (namely, the uncertainty principle and quantum steering) \cite{OW10}.

In addition, it has been recently conjectured \cite{FSABCLA12,Cabello13} that the Tsirelson bound may also be the maximum value allowed by the exclusivity (E) principle, namely, that the sum of probabilities of pairwise exclusive propositions cannot exceed~1. The E principle is a combination of a basic axiom of probability, namely, that the sum of probabilities of {\em jointly} exclusive events cannot exceed~1 \cite{Boole62}, with the observation that, in classical physics and in QM, but not necessarily in other theories \cite{Specker60}, when several propositions are {\em pairwise} decidable, then they are jointly decidable. Specker conjectured that this may be the fundamental principle of QM \cite{Specker09}.

The E principle can be applied to any theory that gives probabilities of {\em events}, i.e., of propositions of the type ``the result $a$ is obtained when test $x$ is performed and the result $b$ is obtained when a test $y$ is performed''; the probability of this event will be denoted as $p(a, b \,|\, x, y)$. The theory also tells when two events are exclusive. It is important to note that we do {\em not} assume that the theory assigns any specific values to the probabilities.

In a recent paper \cite{Yan13}, Yan claims that the E principle, by itself, singles out the maximum quantum value of {\em any} noncontextuality (NC) inequality and, in particular, singles out the Tsirelson bound of the CHSH inequality. However, the proof in Ref.~\cite{Yan13} relies on two assumptions that are not true in general, one assumption that is wrong since it is based on an incorrect use of the E principle, and one extra hidden assumption. This is discussed at the end of this paper. The point is that Yan's result cannot be interpreted as that the E principle, by itself, singles out the Tsirelson bound. However, Yan's paper is probably one of the most important recent contributions for understanding QM, since it is a step in the right direction. In particular, Yan's approach suggests a method to {\em experimentally test} that a violation of the CHSH inequality beyond the Tsirelson bound is impossible, assuming that the E principle holds. The aim of this paper is to describe such an experiment.

\begin{figure}[t]
\begin{center}
\centerline{\includegraphics[scale=0.45]{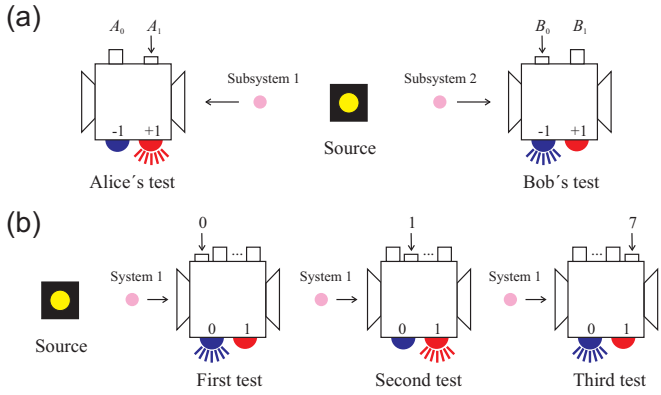}}
\caption{\label{Fig1}(a) Experiment in London to test the CHSH inequality given by (\ref{London}). (b) Experiment in Paris to test the NC inequality given by (\ref{Paris}).}
\end{center}
\end{figure}

{\em A tale of two cities.---}Consider an experiment in London (L) to test the maximum quantum violation of the CHSH inequality, namely,
\begin{equation}
 S_{\rm L} = \sum_{i,j=0}^1 \sum_{a,b} p_{\rm L}(a,b\,|\,i,j) \stackrel{\mbox{\tiny{ LHV}}}{\leq} 3,
 \label{London}
\end{equation}
where the second sum is extended to $a,b \in \{-1,1\}$ satisfying $\frac{1}{2}(a+1)\oplus\frac{1}{2}(b+1)=i j$, where $\oplus$ is addition modulo 2, and
$\stackrel{\mbox{\tiny{ LHV}}}{\leq} 3$ means that the maximum value of $S_{\rm L}$ for local hidden variable theories is $3$. See Fig.~\ref{Fig1}(a).

Suppose that, in this experiment, it is observed that, as predicted by QM \cite{CHSH69}, each of the 8 probabilities in (\ref{London}) is
\begin{equation}
 p_{\rm L} = \frac{2+\sqrt{2}}{8} \approx 0.4267,
 \label{LondonProb}
\end{equation}
thus
\begin{equation}
 S_{\rm L}=2+\sqrt{2} \approx 3.4142,
 \label{LondonMax}
\end{equation}
which is the Tsirelson bound of QM. Suppose that, after obtaining these results, the scientists in London wonder why higher values of $S_{\rm L}$ cannot be reached.


Now consider a second, completely independent, experiment performed in Paris (P). The aim of this second experiment is to test the maximum violation predicted by QM of the following noncontextuality inequality:
\begin{equation}
 S_{\rm P}=\sum_{i=0}^7 p_{\rm P} (0,0,1\,|\,i,i\oplus1,i\oplus2) \stackrel{\mbox{\tiny{ NCHV}}}{\leq} 2,
 \label{Paris}
\end{equation}
where here $\oplus$ means sum modulo 8 and $\stackrel{\mbox{\tiny{ NCHV}}}{\leq} 2$ means that the maximum value of $S_{\rm P}$ for noncontextual hidden variable theories is $2$, something that can be proven using the techniques in \cite{CSW10}. See Fig.~\ref{Fig1}(b).

It can be proven using the techniques in \cite{CSW10} that the maximum possible value of $S_{\rm P}$ in QM is
\begin{equation}
 S_{\rm P}= 8-4\sqrt{2} \approx 2.3431.
 \label{ParisMax}
\end{equation}
As described in the next section, this maximum value is achieved when the value of each of the 8 probabilities in (\ref{Paris}) is
\begin{equation}
 p_{\rm P} = 1-\frac{1}{\sqrt{2}} \approx 0.2929.
 \label{ParisProb}
\end{equation}
Suppose that, after obtaining these results, scientists in Paris wonder why higher values of $S_{\rm P}$ cannot be reached.

Now consider both experiments together. The key point is that the exclusivity graph of $S_{\rm P}$ (i.e., the graph in which events are represented by vertices and exclusive events by adjacent vertices \cite{CSW10}) is the complement (i.e., the one obtained when adjacent vertices become nonadjacent and nonadjacent vertices become adjacent) of the exclusivity graph of $S_{\rm L}$. This implies that we can find sets of 8~pairwise exclusive events by considering both experiments together. The E principle states that the sum of their joint probabilities cannot exceed~1. For example, the following sum of joint probabilities of 8 global events $a, b; c, d, e \,|\, u, v; x, y, z$, defined taking one event $a, b \,|\, u, v$ in London and one event $c, d, e \,|\, x, y, z$ in Paris:
\begin{widetext}
\begin{equation}
\begin{split}
 S_{\rm LP}=&p(1,1; 0,0,1\,|\,0,0;6,7,0)+p(-1,-1; 0,0,1\,|\,0,0;2,3,4)+p(1,1; 0,0,1\,|\,0,1;5,6,7)+p(-1,-1; 0,0,1\,|\,0,1;1,2,3)\\
 &+p(1,1; 0,0,1\,|\,1,0;7,0,1)+p(-1,-1; 0,0,1\,|\,1,0;3,4,5)+p(1,-1; 0,0,1\,|\,1,1;0,1,2)+p(-1,1; 0,0,1\,|\,1,1;4,5,6)
 \label{global}
\end{split}
\end{equation}
\end{widetext}
cannot exceed~1 according to the E principle, since all events in (\ref{global}) are pairwise exclusive. For example, $1,1; 0,0,1\,|\,0,0;6,7,0$ and $-1,1; 0,0,1\,|\,1,1;4,5,6$ are exclusive because test $6$ in Paris has different results.

However, $S_{\rm LP}=1$ is exactly what we obtain when we take into account the results in London (\ref{LondonProb}) and Paris (\ref{ParisProb}), and we take into account that the experiments in London and Paris are independent so
\begin{equation}
 p(a, b; c, d, e \,|\, u, v; x, y, z)=p_{\rm L} (a, b\,|\, u, v) p_{\rm P}(c, d, e \,|\,x, y, z).
\end{equation}
In other words, and assuming that the 8 probabilities $p_{\rm L}$ in London (and $p_{\rm P}$ in Paris) are equal (as will be the case in our experiments), the experiment in London cannot give a higher value for $S_{\rm L}$ without violating the E principle when Paris results are taken into account. Reciprocally, the experiment in Paris cannot give a higher value for $S_{\rm P}$ without violating the E principle when London results are taken into account.

The point is that, while the maximum value of $S_{\rm L}$ predicted by QM has been experimentally observed many times, no experiment so far has tested the maximum value for $S_{\rm P}$ predicted by QM.


{\em How to perform the complementary experiment.---}Here we describe an experiment reaching the maximum value for $S_{\rm P}$ predicted by QM.

Consider a large set of five-dimensional quantum systems prepared in the quantum state $|\psi\rangle$ given by
\begin{equation}
 \langle \psi | = \left(\sqrt{1-\frac{1}{\sqrt{2}}},\sqrt{1-\frac{1}{\sqrt{2}}},\sqrt{1-\frac{1}{\sqrt{2}}},\sqrt{\frac{3}{\sqrt{2}}-2},0\right).
\end{equation}
On each of these systems, we sequentially measure three mutually compatible observables of one of the following sets of observables: $\{0,1,2\}$, $\{1,2,3\}$, $\{2,3,4\}$, $\{3,4,5\}$, $\{4,5,6\}$, $\{5,6,7\}$, $\{6,7,0\}$, and $\{7,0,1\}$. For each set, each of the 6 possible orderings must be tested (since the observables are compatible, the 6 probabilities should be identical in an ideal experiment). The 8 observables in $S_{\rm P}$ are defined as
\begin{equation}
 i = |v_i\rangle \langle v_i|,
 \label{observables}
\end{equation}
with $i=0,\ldots,7$, and
\begin{equation}
\begin{split}
 \langle v_0 |&=(1,0,0,0,0),\\
 \langle v_1 |&=(0,1,0,0,0),\\
 \langle v_2 |&=(0,0,1,0,0),\\
 \langle v_3 |&=\left(2-\sqrt{2},0,0,\sqrt{\sqrt{2}-1},-\sqrt{3 \sqrt{2}-4}\right),\\
 \langle v_4 |&=\left(3-2 \sqrt{2},2-\sqrt{2},0,\sqrt{2 \left(5 \sqrt{2}-7\right)},\sqrt{6 \sqrt{2}-8}\right),\\
 \langle v_5 |&=\left(2-\sqrt{2},3-2 \sqrt{2},2-\sqrt{2},-2 \sqrt{5 \sqrt{2}-7},0\right),\\
 \langle v_6 |&=\left(0,\sqrt{2}-2,2 \sqrt{2}-3,-\sqrt{2 \left(5 \sqrt{2}-7\right)},\sqrt{6 \sqrt{2}-8}\right),\\
 \langle v_7 |&=\left(0,0,\sqrt{2}-2,-\sqrt{\sqrt{2}-1},-\sqrt{3 \sqrt{2}-4}\right).
\end{split}
 \label{OR}
\end{equation}
Each observable $i$ has two possible results: 0 and 1. Observables $i$, $i \oplus 1$, and $i \oplus 2$ (where $\oplus$ is sum modulo 8) are compatible, as required for the NC inequality (\ref{Paris}), since $|v_i\rangle$, $|v_{i \oplus 1}\rangle$, and $|v_{i \oplus 2}\rangle$ are mutually orthogonal. When these observables are measured sequentially on the same system initially prepared in the state $|\psi\rangle$, they exactly reach (\ref{ParisProb}) and (\ref{ParisMax}), namely, the maximum quantum violation of the inequality (\ref{Paris}). 

The proposed experiment is similar to those performed in \cite{KZGKGCBR09,ARBC09,ZUZAWDSDK13,DHANBSC13}, but on a quantum system of dimension five. It can be proven that no quantum system of dimension smaller than five violates inequality (\ref{Paris}) up to its quantum maximum.

It may happen that a simpler NC inequality (or even a Bell inequality) with the same exclusivity graph and violated by QM up to the same maximum
exists. Then, it can be used, instead of inequality (\ref{Paris}), for the argument. 

Notice that the compatibility graph (i.e., the graph in which observables in the NC inequality are represented by vertices and compatible observables are represented by adjacent vertices) is isomorphic to the exclusivity graph for inequality (\ref{Paris}) but not for inequality (\ref{London}).


{\em Discussion.---}The E principle, by itself, {\em exactly} singles out the maximum quantum violation of an NC inequality in three cases: (i) For NC inequalities whose exclusivity graph $G$ is self-complementary, vertex-transitive, and such that the maximum quantum violation equals the Lov\'asz number \cite{Lovasz79} of $G$, $\vartheta(G)$ \cite{Cabello13}, (ii) for fully contextual quantum correlations (defined in \cite{ADLPBC12}), and (iii) for NC inequalities whose exclusivity graph $G$ is vertex-transitive, QM reaches $\vartheta(G)$, and its complement $\bar{G}$ is the exclusivity graph of fully contextual quantum correlations \cite{Cabello13b}. However, in this case, the maximum quantum value equals the maximum allowed by noncontextual theories.

In addition, numerical evidence suggests that the E principle, by itself, applied to an infinite number of copies of the experiment we are interested in, may also single out the maximum quantum violation of the CHSH inequality \cite{FSABCLA12,Cabello13} and the maximum quantum violation of NC inequalities with exclusivity graphs represented by odd cycles on $n$ vertices (with $n$ odd $\ge 7$) and their complements \cite{CDLP12}.

In a recent paper \cite{Yan13}, Yan claims that the E principle, by itself, singles out the maximum quantum value of {\em any} NC inequality. Yan's argument is based on the following four assumptions: (I) The maximum quantum violation of any NC inequality is given by $\vartheta(G)$ of the $n$-vertex exclusivity graph $G$ of the events appearing when the operator in the NC inequality is expressed as a conical sum of probabilities of events $S=\sum_{i=0}^{n-1} p_i(e_i)$, where $e_i$ denotes an event. (II) For every $G$, there is a complementary graph $\bar{G}$ to which one can associate unit vectors $\{|v_i\rangle\}_{i=0}^{n-1}$ to the vertices, such that adjacent vertices are associated orthogonal vectors and such that there is a state $|\psi\rangle$ such that $\sum_{i=0}^{n-1} |\langle \psi | v_i \rangle|^2$ equals $\vartheta(\bar{G})$. (III) For any $G$, the E principle and $\vartheta(\bar{G})$ singles out $\vartheta(G)$. (IV) The vectors $\{|v_i\rangle\}$ in (II) can always be interpreted as events of a certain experiment performed on the state $|\psi\rangle$ such that their exclusivity graph is exactly $\bar{G}$. Then, higher-than-quantum violations of the NC inequality can be excluded by invoking the E principle because, if we define $p_i(\bar{e_i})=|\langle \psi | v_i\rangle|^2$, then the exclusivity graph of the global events $e_i \bar{e_i}$ is the complete graph on $n$ vertices for which the E principle imposes that $\sum_{i=0}^{n-1} p(e_i \bar{e_i})$ cannot exceed~1. Then, since $p(e_i \bar{e_i})=p(e_i) p(\bar{e_i})$, the the E principle prevents $\sum_{i=0}^{n-1} p(e_i)$ to be higher than $\vartheta(G)$.

The problem is that assumption (I) is not true in general (see Ref.~\cite{SBBC13} for counterexamples), assumption (III) is not true in general (see Ref.~\cite{ADLPBC12} for a counterexample), and assumption (IV) is wrong: $\bar{e_i}$, as defined in Yan's argument, {\em is not an event}. Therefore, $e_i \bar{e_i}$ is not an event and then one cannot invoke the E principle which is about the sum of probabilities of events. The main problem arises in fixing assumption (IV), because finding an experiment with events $\{\bar{e_i}\}_{i=0}^{n-1}$ such that their relationships of {\em exclusivity} are represented by an arbitrary graph ($\bar{G}$ in this case) and such that its maximum quantum violation reaches $\vartheta(G)$ is difficult and it is not clear that can it be done for any graph, even when assumptions (I) and (III) hold. For example, it seems to be impossible for the graph in Ref.~\cite{Cabello13c}. Moreover, the proof seems to need an extra assumption, namely, (V) that the maximum is achieved when all the probabilities $p_{\rm L}$ and $p_{\rm P}$ are equal.

Despite all that, the interesting point is that (I) and (III) hold true for the CHSH inequality. Moreover, here we have shown that, for the CHSH inequality, we can fix the problem with assumption (IV). Specifically, we have shown that there is an {\em NC inequality} whose {\em exclusivity graph} is $\bar{G}$ and such that {\em nature violates it up to the Lov\'asz number of} $\bar{G}$. The experimental confirmation of this prediction would explain, assuming the E principle, why higher violations of the CHSH inequality cannot occur.

An open problem is for which NC inequalities such that their maximum quantum violation reaches the Lov\'asz number of the corresponding exclusivity graph $G$ and such that $G$ satisfies (III), there is an NC inequality such that its exclusivity graph is represented by $\bar{G}$ and such that its maximum quantum violation reaches the Lov\'asz number of $\bar{G}$. This can be probably done for many interesting NC inequalities and, specifically, for some important NC inequalities whose exclusivity graph is vertex-transitive.

For example, this can be done for two important families of NC inequalities whose exclusivity graphs are odd cycles on $n$ vertices and their complements \cite{CDLP12}. For these cases, an NC inequality with exclusivity graph $C_n$, a quantum state, a set of observables leading to a violation equal to $\vartheta(C_n)$, a second NC inequality with exclusivity graph $\bar{C_n}$, a second quantum state, and a second set of observables leading to a violation equal to $\vartheta(\bar{C_n})$ are known for any $n$ odd $\ge 7$ \cite{CDLP12}. However, to my knowledge, no experiment testing either of these predictions has yet been performed.

Although Yan's statement in \cite{Yan13} relies on assumptions that are not true in general and one that is wrong, his approach can be used to find experimental evidence that, assuming that the E principle is satisfied, no violation of the CHSH inequality (and other important inequalities) can occur beyond the one predicted by QM.


\begin{acknowledgments}
 I thank M. Ara\'ujo, C. Budroni, O. G{\"u}hne, M. Kleinmann, J.-{\AA}. Larsson,
 A.J. L\'opez-Tarrida, J.R. Portillo, M. Terra Cunha, and B. Yan for useful conversations. This work was supported by the Project No.\ FIS2011-29400 (MINECO, Spain).
\end{acknowledgments}



\end{document}